\documentclass{natcom}

\usepackage{epsfig}
\usepackage{color}
\usepackage{bm}
\usepackage{graphicx}
\usepackage{longtable}
\usepackage{amssymb}
\usepackage{rotating}

\newcommand{\apj}{Astrophys. J.}

\newcommand{\pasp}{Publ. Astron. Soc. Pac.}
\newcommand{\apjs}{Astrophys. J. Supp.}

\newcommand{\mnras}{Mon. Not. R. Astron. Soc.}
\newcommand{\apjl}{Astrophys. J. Let.}
\newcommand{\aap}{Astron. Astrophys.}
\newcommand{\aj}{Astron. J.}
\newcommand{\nat}{Nature}

\def\kms{${\rm km~s^{-1}}$}

\def\DPA{$\Delta {\rm PA}$}
\def\DPAA{\Delta{\rm PA}}

\newcommand{\bc}{\begin{center}}
\newcommand{\ec}{\end{center}}

\title{The growth of the central region
  by acquisition of counter-rotating gas in star-forming galaxies}

\author{Yan-Mei Chen$^{1,2,3,*}$, 
Yong Shi$^{1,2,3}$, 
Christy A. Tremonti$^{4}$, 
Matt Bershady$^{4}$, 
Michael Merrifield$^{5}$, 
Eric Emsellem$^{6,7}$, 
Yi-Fei Jin$^{1,2,3}$, 
Song Huang$^{8}$, 
Hai Fu$^{9}$, 
David A. Wake$^{10}$, 
Kevin Bundy$^{8}$, 
David Stark$^{8}$, 
Lihwai Lin$^{11}$, 
Maria Argudo-Fernandez$^{12,13}$, 
Thaisa Storchi Bergmann$^{14,15}$, 
Dmitry Bizyaev$^{16,17}$, 
Joel Brownstein$^{18}$, 
Martin Bureau$^{19}$, 
John Chisholm$^{4}$, 
Niv Drory$^{20}$, 
Qi Guo$^{21}$, 
Lei Hao$^{12}$, 
Jian Hu$^{22,23}$, 
Cheng Li$^{22,23}$, 
Ran Li$^{21}$, 
Alexandre Roman Lopes$^{24}$,
Kai-Ke Pan$^{16}$, 
Rogemar A. Riffel$^{15, 25}$, 
Daniel Thomas$^{26}$, 
Lan Wang$^{21}$, 
Kyle Westfall$^{26}$, 
Ren-Bin Yan$^{27}$
}

\begin{document}

\maketitle

\let\thefootnote\relax\footnote{
\begin{affiliations}

  \item School of Astronomy and Space Science, Nanjing University, Nanjing 210093, China, chenym@nju.edu.cn
  \item Key Laboratory of Modern Astronomy and Astrophysics (Nanjing University), Ministry of Education, Nanjing 210093, China.\\
  \item Collaborative Innovation Center of Modern Astronomy and Space Exploration, Nanjing 210093, China.\\
  \item Department of Astronomy, University of Wisconsin-Madison, 1150 University Ave, Madison, WI 53706, USA\\ 
  \item School of Physics and Astronomy, University of Nottingham, University Park, Nottingham NG7 2RD, UK\\
  \item European Southern Observatory, Karl-Schwarzschild-Strasse 2, D-85748 Garching, Germany\\
  \item Universit\'e Lyon 1, Observatoire de Lyon, Centre de
  Recherche Astrophysique de Lyon and Ecole Normale Sup\'erieure de Lyon,
  9 avenue Charles Andr\'e, F-69230 Saint-Genis Laval, France\\
  \item Kavli  Institute for the  Physics and Mathematics of
  the Universe (Kavli IPMU, WPI), Todai Institutes for Advanced Study,
  the    University    of     Tokyo,    Kashiwa 277-8583,    Japan\\
  \item Department  of  Physics  and  Astronomy, University
  of Iowa, Iowa  City, IA 52242, USA\\
  \item Department of Physical Sciences,
   The Open University, Milton Keynes, MK7 6AA, UK\\
  \item Institute  of Astronomy  and  Astrophysics, Academia
  Sinica,   Taipei   106,   Taiwan\\
  \item Shanghai Astronomical  Observatory,  Nandan Road  80,
  Shanghai  200030, China\\
  \item Universidad de Antofagasta, Unidad de Astronom'a, Facultad Cs. B‡sicas, Av. U. de Antofagasta 02800, Antofagasta, Chile\\
  \item Departamento de Astronomia, Instituto de F\'\i sica, Universidade Federal do Rio Grande do Sul, CP 15051, 91501-970, Porto Alegre, RS, Brazil\\  
  \item Laborat\'orio Interinstitucional de e-Astronomia - LIneA, Rua Gal. Jos\'e Cristino 77, Rio de Janeiro, RJ - 20921-400, Brazil\\
  \item Apache Point Observatory and New Mexico  State
  University, P.O.  Box 59, Sunspot, NM,  88349-0059,  USA\\
  \item Sternberg Astronomical Institute, Moscow State University, Moscow, Russia\\
  \item Department of Physics and Astronomy, University of Utah, Salt Lake City, UT 84112, USA\\
  \item Sub-Department of Astrophysics, University of Oxford, Denys Wilkinson Building, Keble Road, Oxford OX1 3RH\\
  \item Department of Astronomy and Astrophysics, University
  of California, Santa  Cruz, CA 95064, USA\\
  \item National Astronomical Observatories, Chinese Academy of Sciences, 20A Datun Road, Chaoyang, Beijing 10012, China\\
  \item Department of Physics, Tsinghua University, Beijing 100084, China \\
  \item Center for Astrophysics, Tsinghua University, Beijing 100084, China\\ 
  \item Departamento de Fõsica, Facultad de Ciencias, Universidad de La Serena, Cisternas 1200, La Serena, Chile\\
  \item Departamento de F\'\i sica, Centro de Ci\^encias Naturais e Exatas, Universidade Federal de Santa Maria, 97105-900, Santa Maria, RS, Brazil \\
  \item Institute for Cosmology and Gravitation, University of Portsmouth, Dennis Sciama Building, Burnaby Road, Portsmouth PO1 3FX\\
  \item Department of Physics and Astronomy, University of Kentucky, 505 Rose Street, Lexington, KY 40506-0055, USA\\
\end{affiliations}
}
\vspace{3.5mm}

{\large\bf Abstract}  

Galaxies grow through both internal and external processes. In about 10\% 
of nearby red galaxies with little star formation, gas and
stars are counter-rotating, demonstrating the importance of external
gas acquisition in these galaxies. However, systematic studies of such phenomena in blue,
star-forming galaxies are rare, leaving
uncertain the role of external gas acquisition in driving evolution of
blue galaxies. Based on new measurements with integral field
spectroscopy of a large representative galaxy sample, we find an
appreciable fraction of counter-rotators among blue galaxies (9 out of
489 galaxies). The central regions of blue counter-rotators show
younger stellar populations and more intense, ongoing star formation
than their outer parts, indicating ongoing growth of the central regions. The result
offers observational evidence that the acquisition of external gas in blue
galaxies is possible; the interaction with pre-existing gas funnels
the gas into nuclear regions ($<$ 1 kpc) to form new
stars.  

{\large\bf Introduction}  

In the framework  of hierarchical structure formation,  a galaxy grows
from primordial  density fluctuations and its  subsequent evolution is
shaped by  a series of  external and internal processes.   
Galaxies with gas  and stars  counter-rotating are  the key demonstrations for  the
regulation by external processes\cite{Rubin94, Schweizer98}. External processes, 
e.g. major mergers, minor mergers or gas accretion, could
bring gas which is counter-rotating with pre-existing stars into the galaxies. On the other hand, 
the gas produced by internal processes such as stellar evolution
would conserve the angular  momentum of stars and be co-rotating with pre-existing stars. 

Phenomenon of gas and star  counter-rotating is now known  to be
ubiquitous  in elliptical  and lenticular  galaxies. Still,
the incidence of  gas-star counter-rotators in blue star forming  galaxies is largely
unknown. Since  the early discoveries of  
individual cases\cite{Galletta87},  systematic studies
with long-slit spectroscopy have reported a fraction as high as 
25\%\cite{Bertola92, Kuijken96, Kannappan01} in early type galaxies,  which  decreased to 
a value of 10\%$\sim$15\% with  integral-field
spectroscopy\cite{Sarzi06, Davis11, Barrera15}.  While a few individual cases of blue counter-rotators
are found\cite{Ciri95, Bertola96, Thakar97, Zeeuw02}, existing
statistical studies of blue galaxies failed to identify any blue counter-rotators due
to limited sample size\cite{Kannappan01, Pizzella04} and
instrumentation (e.g., the limited ability of long-slit spectroscopy
to effectively identify the pattern of the star-gas counter-rotating
out of complicated kinematics, particularly in barred
spirals\cite{Galletta96}). 

To place much stronger constraints on the incidence of blue counter-rotators and to understanding the 
influence of gas accretion on the evolution of blue star forming galaxies, in this work we study 
a sample of galaxies observed with fiber-optic integral-field units
(IFU) in the first year of the survey: Mapping Nearby Galaxies at Apache Point Observatory (MaNGA)\cite{Bundy15}, finding 
$\sim$2\% blue star-forming galaxies have counter-rotating gas.
The central regions of blue counter-rotators show younger stellar populations and more intense, ongoing 
star formation than their outer-skirts, indicating that these galaxies accrete abundant external gas, 
the interaction with pre-existing gas triggers the gas into central regions, and form new
stars.

{\large\bf Results} 

\begin{figure*}
\centerline{ \includegraphics[width=0.8\textwidth]{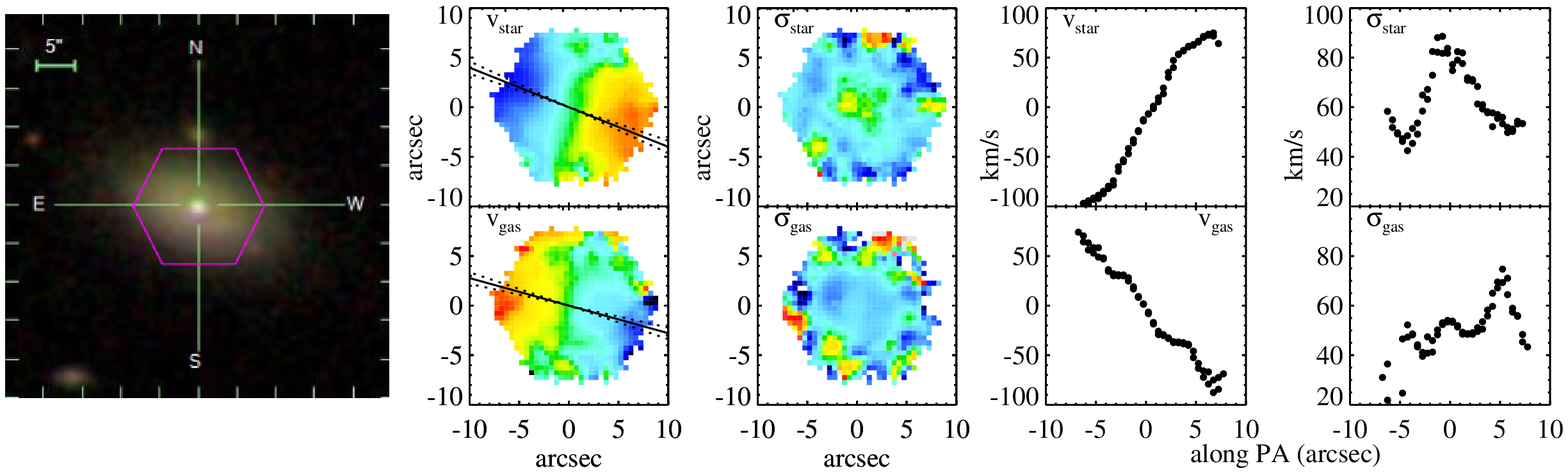} }
\vspace{-4mm}
\caption{{\bf An example of a blue star-forming counter-rotating
  galaxy.} The left panel shows the SDSS $g,r,i-$band image, the projected
  velocity fields of stars (top) and gas (bottom) are shown in the
  second column, while the third column shows the velocity dispersion
  maps of stars and gas. The projected velocity and velocity
  dispersion along major axis (black solid line in the second column)
  are shown in the last two columns. Dashed black lines represent
  $\pm1\sigma$ uncertainties in the major-axis position angle.}
\vspace{-4mm}
\label{fig:examplect}
\end{figure*}

{\bf Sample selection}\\
We analyze gas and stellar kinematic maps of a representative sample
of 1351 nearby galaxies with stellar masses above 10$^{9}$ solar
mass from MaNGA. Fig.~\ref{fig:examplect} shows an
example of a counter-rotating blue star forming galaxy. The Sloan Digital Sky Survey (SDSS)
false-color image is at left, while the kinematics based on
spectroscopic IFU data for stars and gas are mapped in the second and
third columns (velocities and velocity dispersions, respectively). To
quantify the kinematic misalignment between stars and gas, we measured
the difference in the kinematic position angle (PA) between ionized
gas and stars as $\Delta {\rm PA=|PA_{*}-PA_{\rm gas}}|$, where ${\rm
  PA_{*}}$ is the PA of stars and ${\rm PA_{\rm gas}}$ is the PA of
ionized gas. The kinematic PA is measured based on established
methods\cite{Krajnovic06}, defined as the counter-clockwise angle
between north and a line which bisects the velocity field of gas or
stars, measured on the receding side. The solid lines in Fig.~\ref{fig:examplect} show
the best fit position angle and the two dashed lines show the
1-$\sigma$ error. The last two columns show the rotation velocity and
velocity dispersion along the major axis.

\begin{figure*}
\centerline{ \includegraphics[width=1.0\textwidth]{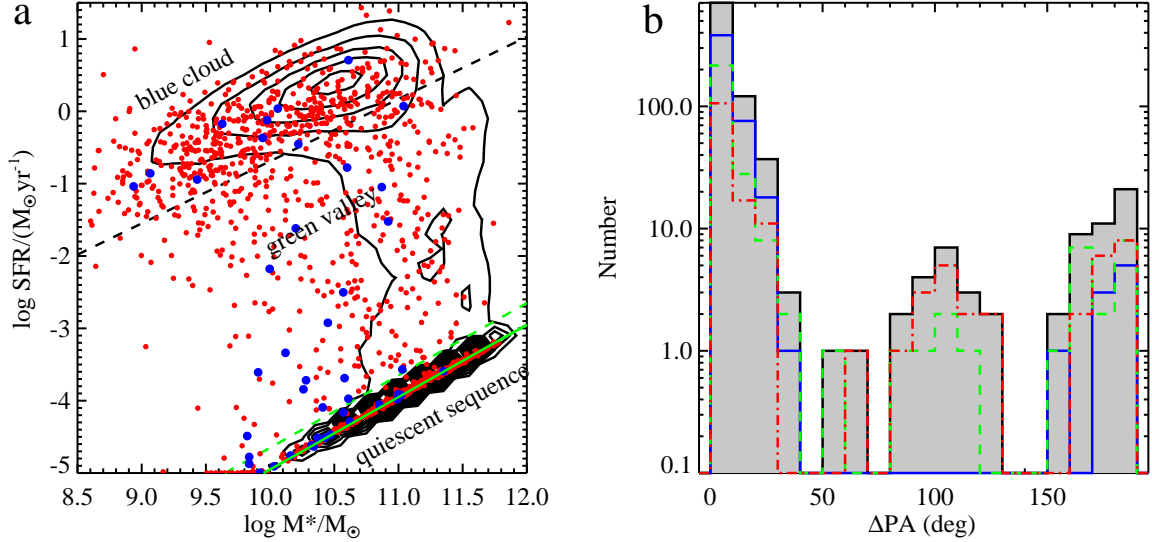} }
\vspace{-4mm}
\caption{{\bf SFRs versus stellar masses and $\Delta$PA distribution.} (a) SFRs vs. stellar mass. Contours show the SDSS DR7
  sample, while the red dots are MaNGA galaxies. The blue dots are the
  counter-rotators with ${\rm \Delta PA}>150^\circ$. The two dashed
  lines separate the galaxies into blue star-formers, green valley,
  and red quiescent galaxies. The black dashed line is adopted from
  Fig.~11 of $ref.$\cite{Chang15} as an approximation of the
  boundary (at the 1$\sigma$ level in scatter) of the star-forming
  main sequence. The green solid line with log sSFR ($\equiv {\rm SFR}/M_*$) $\sim-$15 remarks
  red galaxies in which the SFR can be neglected. The region between
  the black and green dashed lines is referred as the green
  valley. Although galaxies in the green valley have low SFR, they are
  clearly distinguished from red galaxies. We do not use the
  color-magnitude diagram to separate blue from green and red galaxies
  since the colors are strongly effected by dust extinction. (b)
  $\Delta$PA distribution for MaNGA galaxies with nebular
  emission. The grey histogram is for the whole sample, red for the
  red quiescent galaxies.}
\vspace{-4mm}
\label{fig:class}
\end{figure*}

We matched the MaNGA sample with the literature catalog\cite{Chang15}
to obtain the global star formation rate (SFR) and stellar mass ($M_*$)
for 1220 out of 1351 galaxies. With these two quantities we classify
the sample into blue star-forming galaxies, red quiescent galaxies
with little star formation, and green-valley galaxies between these
two extremes (see Fig.~\ref{fig:class}a), as summarized in Table~1. For
simplicity, we refer to these three classes as blue, red and green
galaxies henceforth. Fig.~\ref{fig:class}b shows the distributions of $\Delta {\rm
  PA}$ for these different types of galaxies with nebular emission
(required to measure the gas kinematics). Both green (green histogram) 
and red (red histogram) galaxies have a distribution of the  ${\rm \Delta PA}$, with the three 
local peaks at ${\rm \Delta PA}$ = 0$^\circ$, 90$^\circ$
and 150$^\circ$, while blue galaxies (blue histogram) present a bimodal distribution (the lack of 
a third peak at 90$^\circ$ being consistent with small number statistics). The grey histogram is for the whole population -- the
combination of blue, red and green.  In total there are 43
counter-rotators, i.e.  galaxies with ${\rm \Delta PA}>150^\circ$.
Considering the completeness correction of the MaNGA sample, the
fraction of the counter rotators in blue galaxies is 2\% (9 out of
489), while the fractions in red and green galaxies are 10\% (16 out
of 164) and 6\% (18 out of 280), respectively. Our fraction of
counter-rotators in the red galaxies is consistent with previous
studies\cite{Bertola92, Kuijken96, Sarzi06, Davis11, Krajnovic11,
  Barrera15}.  Thanks to the unbiased MaNGA galaxy sample with respect to morphology,
inclination, color, etc, we can study the incidence as well as the properties of 
blue counter-rotators for the first time. The above fractions could be lower limits,
since for face-on galaxies, it is not possible to measure rotation.

\begin{figure*}
\centerline{ \includegraphics[width=0.8\textwidth]{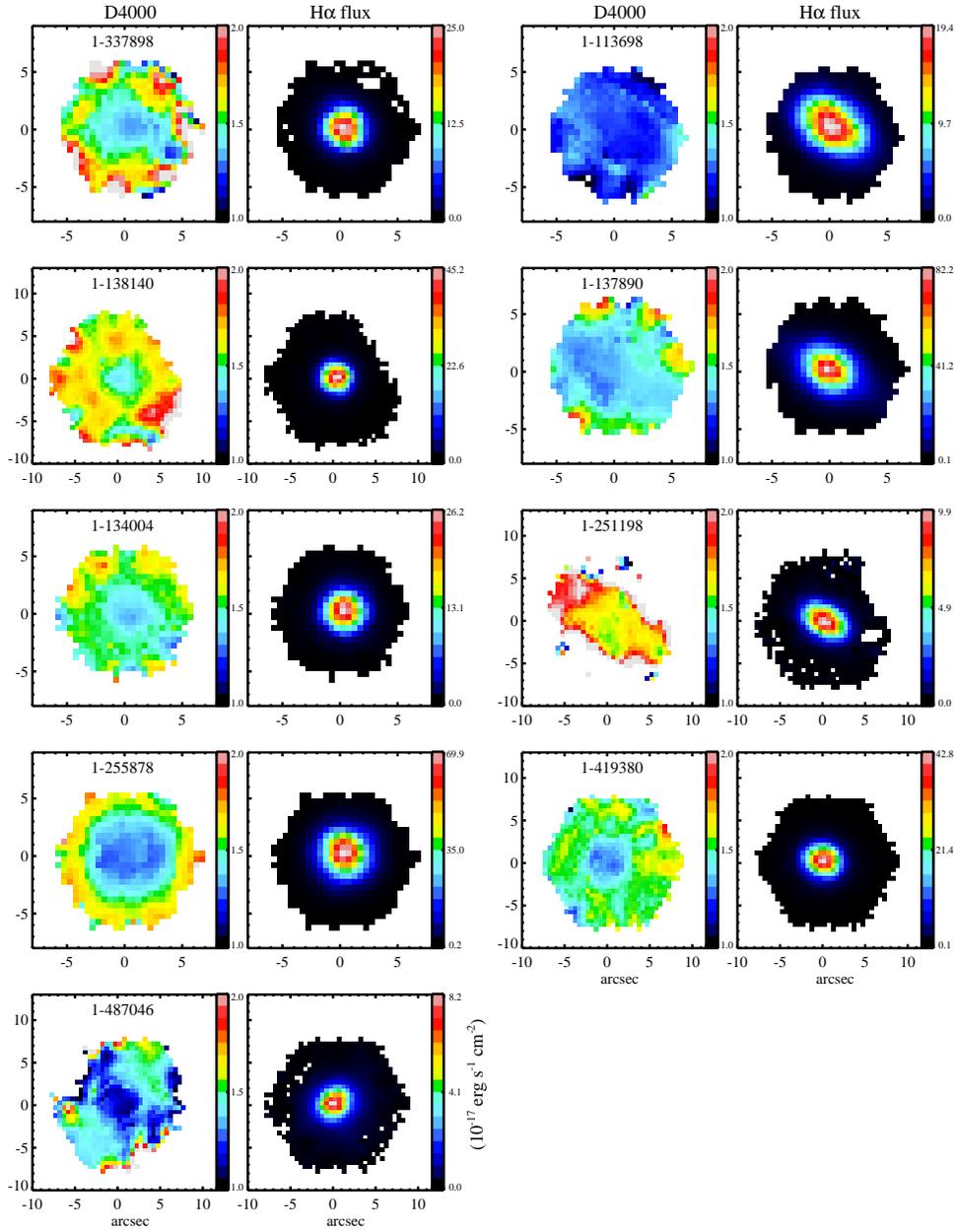} }
\vspace{-4mm}
\caption{{\bf The D4000 and H$\alpha$ flux maps for nine star-forming
  counter rotators.} The MaNGA-ID for each galaxy is shown in the D4000
  map. The H$\alpha$ flux is in the unit of $10^{-17}{\rm erg}~{\rm
    s}^{-1}~{\rm cm}^{-2}$.}
\vspace{-4mm}
\label{fig:d4n}
\end{figure*}

{\bf Properties of blue star-forming counter-rotators}\\
Among nine blue counter-rotators, six of them have strong positive
gradients in the 4000 \AA\ break (D4000) as shown in Fig.~\ref{fig:d4n} while
the remaining show small D4000 across the whole galaxy body,
indicating young stellar populations existing in the central regions.
The map of the H$\alpha$ flux further shows ongoing star formation in
the central region.  We checked the emission line ratio
diagnostic\cite{Baldwin81} to assure that the H$\alpha$ radiation is
dominated by star formation instead of active galactic nuclei (AGN;
see the Fig.~5). In contrast to the blue
counter-rotators, all the green and red counter-rotators have negative
D4000 gradients with older stellar populations in the central
regions. Although the H$\alpha$ flux also peaks at the center for the
green and red counter rotators, it is primarily contributed by the AGN
based on the emission-line diagnostic\cite{Baldwin81}.

\begin{figure*}
\centerline{ \includegraphics[width=0.8\textwidth]{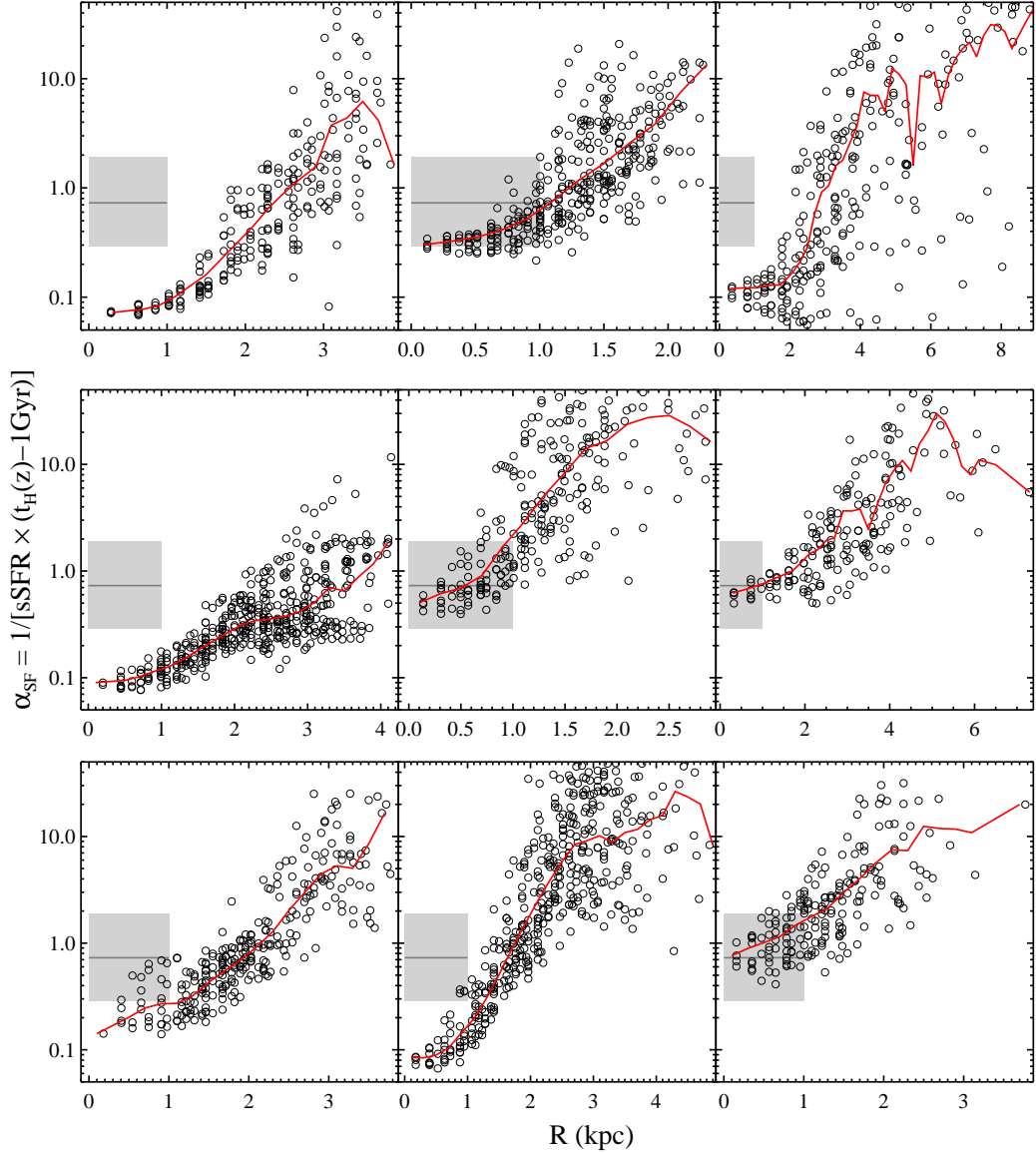} }
\vspace{-4mm}
\caption{{\bf The star formation activity parameter $\alpha_{\rm SF}$ versus radius for the nine blue star forming counter
  rotators.} The circles are our data points while the red lines show
  the median. The grey shaded regions show the $\pm1\sigma$ range of
  $\alpha_{\rm SF}$ for the central 1kpc of local star forming
  galaxies with ${\rm \Delta PA}<30^\circ$.  Grey lines mark the
  median value of $\sim$0.75.} 
\vspace{-4mm}
\label{fig:asf}
\end{figure*}

To further quantify the importance of the ongoing star formation in growing the 
central region, we introduce the star formation activity parameter\cite{Dave08} as 
$\alpha_{\rm SF} = 1/[{\rm sSFR}\times{(t_H(z)-1{\rm Gyr})}]$, where $t_H(z)$ is the 
Hubble time at the redshift of the galaxy, and 1 Gyr is subtracted to account for the 
fact that star formation mainly occurred after reionization. If a galaxy's current 
SFR is equal to its past average ($M_*/((t_H(z)-1{\rm Gyr})$) then $\alpha_{\rm SF}=1$;  values 
less than one indicate that the current SFR is higher than the past average. 
As shown in Fig.~\ref{fig:asf}, all nine galaxies
present a steep rising $\alpha_{\rm SF}$ with increasing distances
from the galaxy center. The grey shaded regions show the $\pm1\sigma$
range of $\alpha_{\rm SF}$ for the central 1kpc of local star forming
galaxies with ${\rm \Delta PA}<30^\circ$.  Grey lines mark the median
value of $\sim$0.75. Focusing on the central 1 kpc, we find six of the
blue counter rotators have $\alpha_{\rm SF}$ about one order of
magnitude smaller than the average value (the grey line), indicating
fast growth of the central components of these galaxies.

Both the D4000 and star formation activity parameter $\alpha_{\rm SF}$ suggest
significant ongoing growth of the central region ($<$ 1 kpc) of these
blue counter-rotators by star formation.   
For nine blue counter-rotators, we fit the $r$-band surface brightness profiles (see
Fig.~5) and found that five of them already
have photometric bulge-like components (above an exponential disk-like component). In addition, the SDSS images show
no signs of strong galaxy interactions or major merging, indicating
accretion of gas from intergalactic medium or dwarfs (minor mergers)
as the origin of the counter-rotating gas. This is also consistent
with their environments, as both the neighbor number ($N$) and the
tidal strength parameter\cite{Argudo15} ($Q_{\rm lss}$) indicate that the
blue counter rotators tend to be located in more isolated
environments. By matching our galaxies with the MPA-JHU catalog
(http://wwwmpa.mpa-garching.mpg.de/SDSS/DR7/oh.html), we obtained the
metallicity for eight blue counter-rotators.  Four of them follow the
stellar mass vs. metallicity relation of the general
population\cite{Tremonti04}, while another four lie 0.2-0.3 dex above
the stellar-mass vs. metallicity relation.

{\large\bf Discussion}

We suggest the following scenario to explain the above observational
facts: (i) The progenitor accretes counter-rotating gas from a
gas-rich dwarf or cosmic web. (ii) Redistribution of angular momentum
occurs from gas$-$gas collisions between the pre-existing and the
accreted gas largely accelerates gas inflow, leading to a fast
centrally-concentrated star formation. (iii) Higher metallicity is a puzzle, one possibility is due
to the enrichment from star formation. In a closed-box
model\cite{Dalcanton07}, the metallicity will mainly depend on the gas
mass fraction $f_{\rm gas}$ ($\equiv M_{\rm gas}/(M_{\rm gas}+M_{\rm stars})$), so the
abundances get elevated instantaneously as a large fraction of the
available gas turns into stars. The low D4000 at the center is a hint
that such stars exist. However, we keep in mind that the `external' gas likely had 
low metallicity and the closed-box model is a strong assumption, future simulations 
are necessary in helping us to understand the gas enrich process.

Though the amount of pre-existing and
accreted gas in the nine galaxies is uncertain, collision between
pre-existing and accreted gas is unavoidable, leading to redistribution
of angular momentum and dissipation of kinetic energy. The impact on 
both the morphology and dynamics of the inner parts of the galaxy may 
thus be associated with the observed slight increase of the gas velocity dispersion.
We find the typical gas velocity
dispersion ($40\sim$ 60\,\kms) in the disk region of these nine
galaxies is about 20 \kms\ larger than a control sample of
star-forming galaxies with aligned gas and stellar kinematics (${\rm
  \Delta PA}<30^\circ$), closely matched in SFR, $M_*$ and redshift. The typical 
errors of gas velocity dispersion is about  10\kms.

In summary, redistribution of angular
momentum through the collisions between accreted and pre-existing gas
is thus an efficient way for gas to migrate to the centre, indicating
that accretion of counter-rotating gas into disk galaxies is an
effective way to grow the central region. This mechanism may be more
effective in growing the central component of galaxies at $z\sim1-2$ where 
external gas acquisition is more frequent\cite{Genzel14, Danovich15}.

{\large\bf Methods}

{\bf Observations and data reduction}\\
The data used in this work comes from the ongoing MaNGA
survey\cite{Bundy15,Drory15,Law15,Yan16} using the SDSS 2.5-in
telescope\cite{Gunn06} and Baryon Oscillation Spectroscopic Survey (BOSS) spectrographs\cite{Smee13}. As one of
three programs comprising the SDSS-IV, MaNGA is
obtaining spatially resolved spectroscopy for about 10,000 nearby galaxies
with log~$M_*/M_\odot \geq$~9 and a median redshift of $z \approx
0.04$. The $r$-band signal-to-noise ratio (S/N) in the outskirts of
MaNGA galaxies is 4$-$8~\AA$^{-1}$, and the wavelength coverage is
3600$-$10300\AA. MaNGA's effective spatial and spectral resolution is
2$^{\prime\prime}$.4 (Full Width at Half Maximum, FWHM) and $\sigma \sim$~60~km~s$^{-1}$,
respectively. The MaNGA sample and data products used here were drawn
from the internal MaNGA Product Launch-4 (MPL-4), which includes
$\approx$ 1400 galaxies observed through July 2015 (the first year of
the survey).

The MaNGA Data Analysis Pipeline (DAP), which uses
pPXF\cite{Cappellari04} and the MIUSCAT stellar
library\cite{Vazdekis12}, fits the stellar continuum in each spaxel
and produces estimates of the stellar kinematics. Ionized gas
kinematics, $v_{\rm gas}$ and $\sigma_{\rm gas}$, as well as the flux
were estimated by fitting a single Gaussian to the emission lines
after stellar continuum subtraction. The observables used in this
work, i.e.  $v_{\rm gas}$ and $\sigma_{\rm gas}$, D4000, emission line
flux, are from DAP.

{\bf Redshift distributions of the samples}\\ 
In Fig.~5, we show the redshift distributions 
of the whole MaNGA sample (black histogram), the blue (blue histogram), red (red histogram) 
and green (green histogram) subsamples as well as the nine blue counter rotators (cyan histogram).

{\bf Sample completeness correction}\\  
An issue with every data set is the selection of weights to correct
for missing galaxies. The MaNGA target sample is selected to lie
within a redshift range, $z_{\rm min} < z < z_{\rm max}$, that depends on
absolute $i-$band magnitude in the case of the Primary and Secondary
samples, and absolute $i-$band magnitude and $NUV-r$ color in the case
of the color-enhanced (CE) sample. $z_{\rm min}$ and $z_{\rm max}$ are chosen
to yield both the same number density of galaxies and angular size
distributions, matched to the IFU sizes, at all absolute $i-$band
magnitudes (or magnitudes and colors for the CE sample). This results
in lower, and narrower, redshift ranges for less luminous galaxies and
higher and wider redshift ranges for more luminous galaxies.

At a given $M_{\rm i}$ (or $M_{\rm i}$ and $NUV-r$ color for the CE sample) the
sample is effectively volume limited in that all galaxies within
$z_{\rm min}(M_{\rm i}) < z < z_{\rm max}(M_{\rm i})$ are targeted irrespective of their
other properties. However, that volume varies with $M_{\rm i}$. Therefore in
any analysis of the properties of MaNGA galaxies as a function of
anything other than $M_{\rm i}$ we must correct for this varying selection
volume, $V_{\rm s}(M_{\rm i})$ -- the volume with $z_{\rm min}(M_{\rm i}) < z <
z_{\rm max}(M_{\rm i})$.  The simplest approach is just to correct the galaxies
back to a volume-limited sample by applying a weight ($W$) to each
galaxy in any calculation such that $W = V_{\rm r}/V_{\rm s}$ where $V_{\rm r}$ is an
arbitrary reference volume. Since the $z_{\rm min}$ and $z_{\rm max}$ for each
MaNGA galaxy are provided in the MaNGA sample catalogue (D.A. Wake et
al. in preparation), we can easily estimate the fraction of galaxies
with decoupled gas and star kinematics in a complete sample by
applying this volume correction.

{\bf Global SFR and $M_*$}\\
Combining SDSS and Wide-field Infrared Survey Explorer (WISE) photometry for the full SDSS spectroscopic
galaxy sample, the spectral energy distributions (SEDs) that cover
$\lambda$ = 0.4 $\sim$ 22 $\mu$m has been created for a sample of
858,365 present-epoch galaxies\cite{Chang15}.  Using MAGPHYS\cite{Cunha08}, they
then model both the attenuated stellar SED and the dust emission at 12
and 22 $\mu$m, producing new calibrations for monochromatic mid-IR SFR
proxies as well as $M_*$.

{\bf Spatially  resolved SFR  and $M_*$} \\
Principal component analysis (PCA) is a standard multivariate analysis
technique, designed to identify correlations in large data sets.
Using PCA, a new method\cite{Chen12} has been generated to estimate
stellar masses, mean stellar ages, star formation histories (SFHs),
dust extinctions and stellar velocity dispersions for galaxies from BOSS.  To obtain these
results, we use the stellar population synthesis models of
BC03\cite{BC03} to generate a library of model spectra with a broad
range of SFHs, metallicities, dust extinctions and stellar velocity
dispersions. The PCA is run on this library to identify its principal
components (PC) over a certain rest-frame wavelength range
3700$-$5500$\rm \AA$.  We then project both the model spectra and the
observed spectra onto the first seven PCs to get the coefficients of
the PCs, which represents the strength of each PC presented in the
model or observed spectra. We derive statistical estimates of various
physical parameters by comparing the projection coefficients of the
observed galaxy to those of the models as follows.  The $\chi^2$
goodness of fit of each model determines the weight $\sim {\rm
exp(-\chi^2/2)}$ to be assigned to the physical parameters of that
model when building the probability distributions of the parameters of
the given galaxy. The probability density function (PDF) of a given
physical parameter is thus obtained from the distribution of the
weights of all models in the library.  We characterize the PDF using
the median and the 16\%$-$84\% range (equivalent to $\pm$1$\sigma$
range for Gaussian distributions). In this work, we directly apply
this PCA method to the MaNGA data to get the stellar mass for each
spaxel.

The SFR for each spaxel is derived from the dereddened H$\alpha$
luminosity ($L_{\rm H\alpha}$) as ${\rm SFR}~(M_\odot~{\rm yr^{-1}}) =
7.9\times10^{-42}~L_{\rm H\alpha}~$(erg~s$^{-1}$).  We use Balmer
decreasement for dust extinction correction.

{\bf Environment} \\
We characterize the environment with two parameters, the neighbor
number ($N$) and the tidal strength parameter $Q_{\rm lss}$. The neighbor
number is defined as the count of galaxies brighter than $-$19.5 mag
in $r$-band absolute magnitude within a fixed volume of 1 Mpc in
projected radius and 500 km s$^{-1}$ in redshift to the primary
galaxy. Given the neighbor number is independent of the stellar mass
and cannot account for the interaction a galaxy suffering from its
satellites, we also use the tidal strength parameter $Q_{\rm lss}$ to
depict the effect of total interaction strength produced by all the
neighbors within the fixed volume\cite{Argudo15, Verley07}; the higher
the parameter, the stronger the interaction. The parameter $Q_{\rm lss}$
is defined as
\begin{equation} \label{equ:qlss}
Q_{\rm lss} \equiv {\rm log}\left[\sum_{i} \frac{M_{i}}{M_{\rm p}}\left(\frac{D_{\rm p}}{d_{i}}\right)^{3}\right]
\end{equation}
where $M_{i}$ and $M_{\rm p}$ are the stellar masses of the $i^{th}$
neighbor and the primary galaxy. $d_{i}$ is the projected distance
from the primary galaxy to the $i^{th}$ satellite and $D_{\rm p}$ is the
estimated diameter of the central galaxy\cite{Argudo15}. Both the
number of neighbors and $Q_{\rm lss}$ are drawn from the catalogue
generated by Argudo-Fern{\'a}ndez et al.

{\bf  Surface brightness profile} \\
We fit the surface brightness profiles of the nine blue counter
rotators with three different models: (1) single Sersic; (2) double
Sersic; (3) Sersic bulge + exponential disk. The best fitting results
are shown Fig.~6 \& 7.

{\bf Data availability}\\ 
The data supporting the findings of this study are available through 
SDSS Data Release Thirteen which can be downloaded from http://www.sdss.org/dr13/manga/.

{\large\bf References}

\bibliographystyle{naturemag,aastex}

{\large\bf Acknowledgements}

We are very grateful to the anonymous referees for useful comments
and suggestions that have strengthened this work. Y.M.C. acknowledges support from NSFC grant 11573013, 11133001, 
the Natural Science Foundation of Jiangsu Province grant BK20131263, the Opening Project of Key 
Laboratory of Computational Astrophysics, National Astronomical Observatories, 
Chinese Academy of Sciences. Y.S. acknowledges support from  NSFC  grant  11373021,  
the CAS Pilot-b grant No. XDB09000000 and Jiangsu Scientific Committee grant BK20150014.
C.A.T. acknowledges support from National Science Foundation of the United States Grant No. 1412287.

Funding for the Sloan Digital Sky Survey IV has been provided by the Alfred P. Sloan Foundation, the U.S. Department of Energy Office of Science, and the Participating Institutions. SDSS- IV acknowledges support and resources from the Center for High-Performance Computing at the University of Utah. The SDSS web site is www.sdss.org.

SDSS-IV is managed by the Astrophysical Research Consortium for the Participating Institutions of the SDSS Collaboration including the Brazilian Participation Group, the Carnegie Institution for Science, Carnegie Mellon University, the Chilean Participation Group, the French Participation Group, Harvard-Smithsonian Center for Astrophysics, Instituto de Astrof\'{i}sica de Canarias, The Johns Hopkins University, Kavli Institute for the Physics and Mathematics of the Universe (IPMU) / University of Tokyo, Lawrence Berkeley National Laboratory, Leibniz Institut f\"{u}r  Astrophysik Potsdam (AIP), Max-Planck-Institut f\"{u}r  Astronomie (MPIA Heidelberg), Max-Planck-Institut f\"{u}r Astrophysik (MPA Garching), Max-Planck-Institut f\"{u}r  Extraterrestrische Physik (MPE), National Astronomical Observatory of China, New Mexico State University, New York University, University of Notre Dame, Observat\'{o}rio Nacional / MCTI, The Ohio State University, Pennsylvania State University, Shanghai Astronomical Observatory, United Kingdom Participation Group, Universidad Nacional Aut\'{o}noma de M\'{e}xico, University of Arizona, University of Colorado Boulder, University of Oxford, University of Portsmouth, University of Utah, University of Virginia, University of Washington, University of Wisconsin, Vanderbilt University, and Yale University.

{\large\bf Author information}

{\bf Author  Contributions} \\
Y.M.C. discovered these sources, studied their properties and led the writing of the manuscript. Y.S., C.A.T., M.B., M.M. and E.E. provided the picture to explain all the observation results. Y.S. and M.M. also helped the writing of the manuscript. Y.F.J. helped in making plots. S.H. fitted the surface brightness profile. D.A.W. helped with the sample completeness correction. D.S. searched for deeper images of these galaxies. L.L and M.A.F. provided the environment parameters.  K.B., R.B.Y., M.B., N.D., D.A.W., D.T. and D.B. contributed to the design and 
execution of the survey. All authors commented on the manuscript and contributed to the interpretation of the observations. 

{\bf Competing interests} \\
The authors  declare no  competing financial interests. 

{\bf Corresponding author}\\
Correspondence and requests for materials should be addressed to Y.M.C. (chenym@nju.edu.cn).

\clearpage

\begin{figure*}
\centerline{ \includegraphics[width=0.8\textwidth]{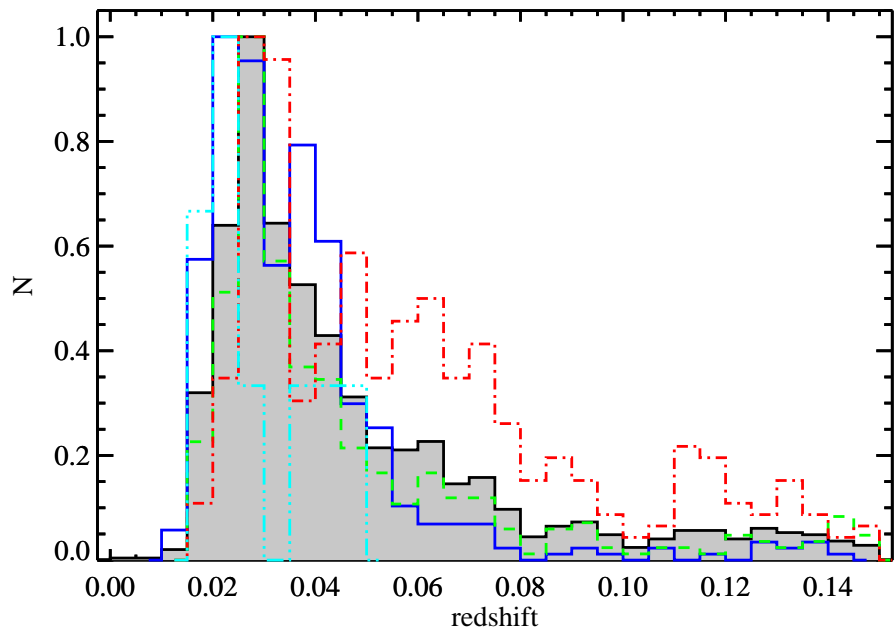} }
\vspace{-4mm}
\caption{{\bf The redshift distributions of the samples.}
The grey histogram is for the whole MaNGA sample; the blue, red  
and green histograms show the redshift distributions for the blue, red and green subsamples, respectively; 
the nine blue counter rotators are shown in cyan histogram.} 
\vspace{-4mm}
\label{fig:zdist}
\end{figure*}

 \begin{figure*}
\centerline{ \includegraphics[width=0.8\textwidth]{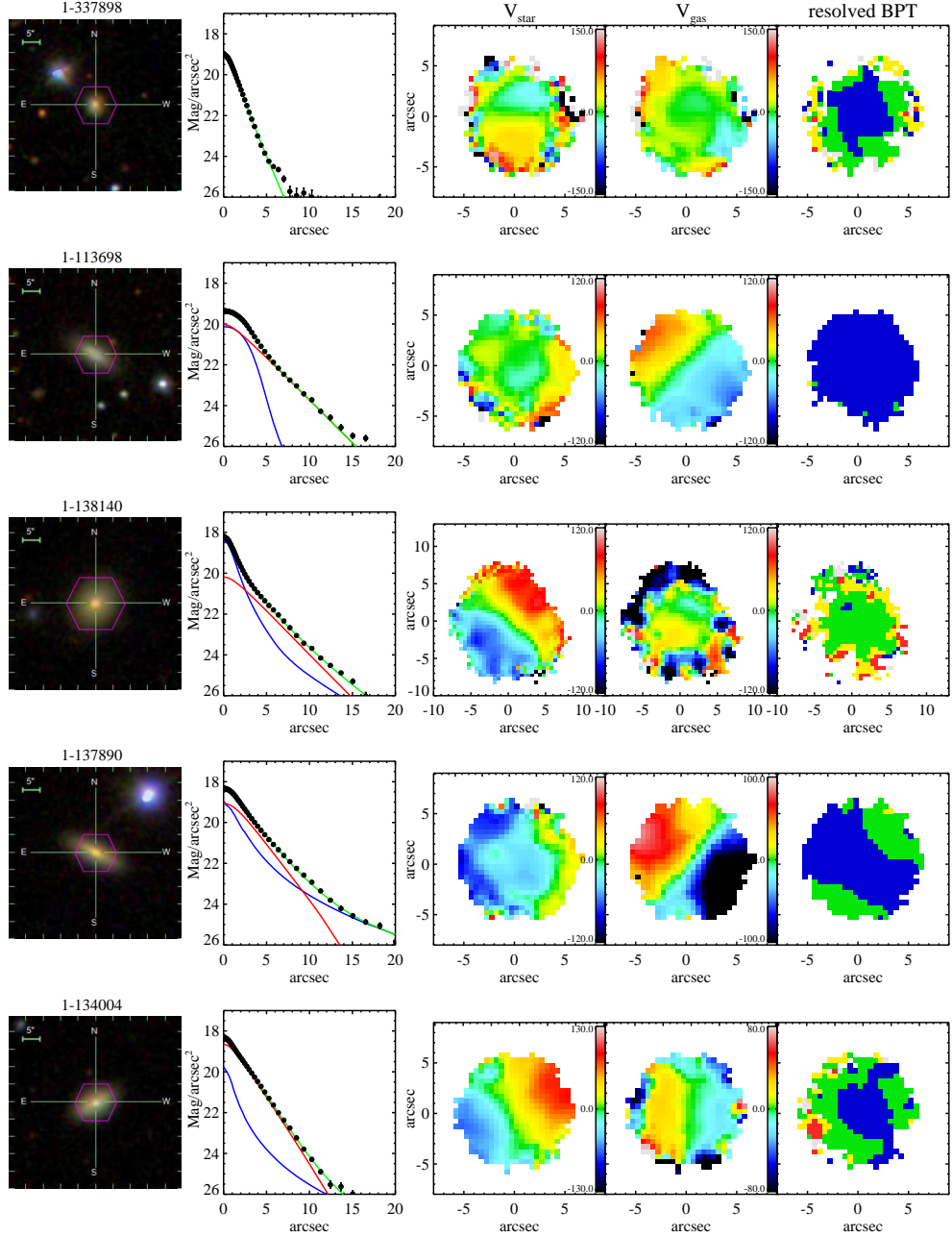} }
\vspace{-4mm}
\caption{{\bf Properties of the blue counter rotators.} Left: the SDSS 
false-color image; second column: the surface brightness profile,
black is the data, green is the best fit model. Except for the first
object, all the others are fitted by two components (red + blue); the
third and fourth columns show the velocity fields of stars and gas,
respectively. The velocities are in the unit of $\rm km~s^{-1}$. The spatial resolved BPT diagram\cite{Baldwin81} is shown in the last
column, blue represents star forming region, red represents Seyfert,
green is the composite of AGN and star formation and yellow represents
Low-Ionization Emission-line Region (LIER).}
\vspace{-4mm}
\end{figure*}

\begin{figure*}
\centerline{ \includegraphics[width=0.8\textwidth]{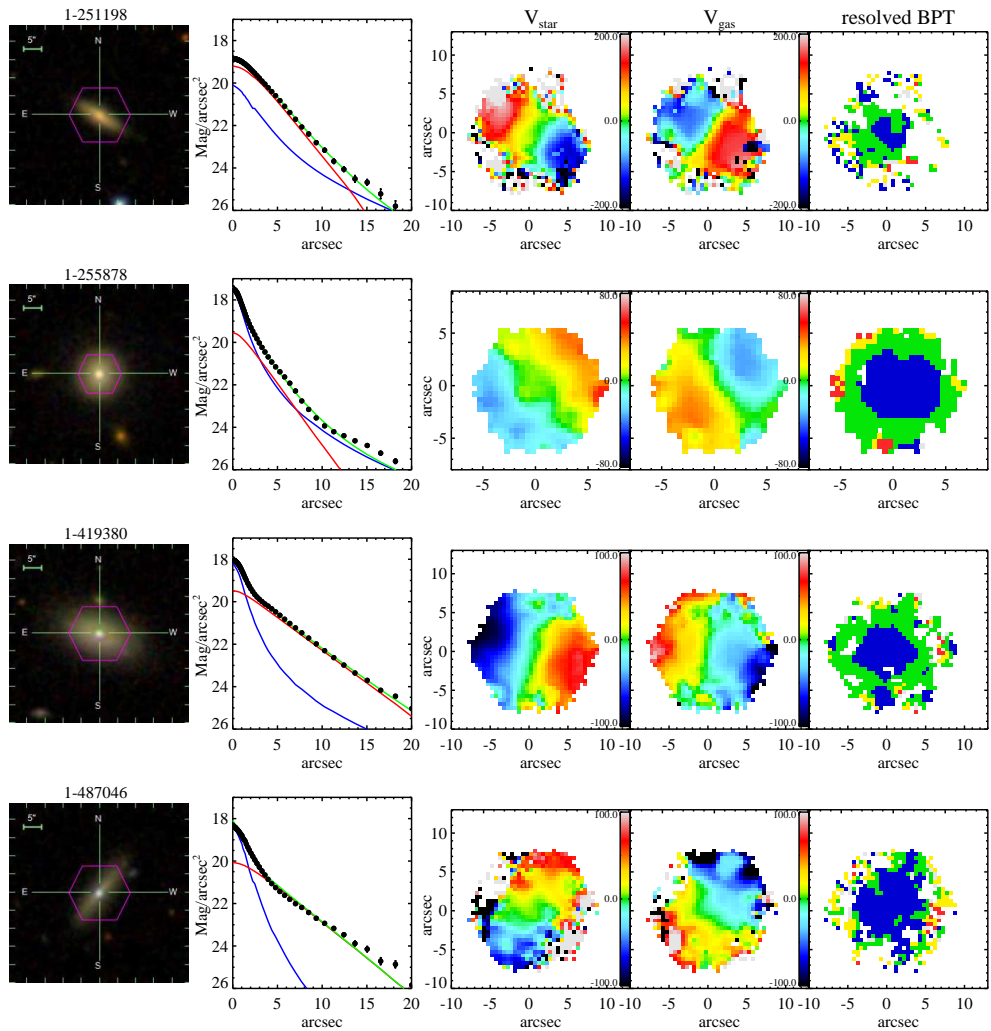} }
\vspace{-4mm}
\caption{{\bf Properties of the blue counter rotators.} Same as Fig.~6, but with more objects.}
\vspace{-4mm}
\end{figure*}

\clearpage

\begin{table*}
\begin{center}
\begin{tabular}{lcccc}\hline\hline
  type & number              & misalignment & & counter-rotators  \\ \cline{3-3}\cline{5-5}
       & (number with EML)   &($\DPAA>30^{\circ}$)& & ($\DPAA>150^{\circ}$)\\ \hline
blue   &489 (489)                   & 10           & &9 \\
green  &377 (280)                   & 26           & &18 \\
red    &354 (164)                   & 30           & &16\\
total  &1220 (933)                  & 66           & &43 \\ \hline
\end{tabular}
\end{center}
{\textbf Table 1 $|$ Classification of the MaNGA sample.}
This table gives the number of galaxies in each catagory.  blue: blue
star forming galaxies; green: green valley; red: red quiescent
galaxies. Misalignment and counter-roators are classified by \DPA\,
given in the table. EML means galaxies with emission lines; the number
of galaxies with line emission is in parenthesis.
\end{table*}

\end{document}